# Representing provenance and track changes of cultural heritage metadata in RDF: a survey of existing approaches

*Arcangelo Massari [1,2], Silvio Peroni [1,2], Francesca Tomasi [2], and Ivan Heibi [1,2]*
*[1] Research Centre for Open Scholarly Metadata, Department of Classical Philology and Italian Studies, Alma Mater Studiorum University of Bologna, Bologna, Italy*
*[2] Digital Humanities Advanced Research Centre (/DH.arc), Department of Classical Philology and Italian Studies, Alma Mater Studiorum University of Bologna, Bologna, Italy*

## Abstract

In the realm of Digital Humanities, the management of cultural heritage metadata is pivotal for ensuring data trustworthiness. Provenance information—contextual metadata detailing the origin and history of data—plays a crucial role in this process. However, tracking provenance and changes in metadata using the Resource Description Framework (RDF) presents significant challenges due to the limitations of foundational Semantic Web technologies. This article offers a comprehensive review of existing models and approaches for representing provenance and tracking changes in RDF, with a specific focus on cultural heritage metadata. It examines W3C standard proposals such as RDF Reification and n-ary relations, along with various alternative systems. Through an in-depth analysis, the study identifies Named Graphs, RDF-star, the Provenance Ontology (PROV-O), Dublin Core (DC), Conjectural Graphs, and the OpenCitations Data Model (OCDM) as the most effective solutions. These models are evaluated based on their compliance with RDF standards, scalability, and applicability across different domains. The findings underscore the importance of selecting the appropriate model to ensure robust and reliable management of provenance in RDF datasets, thereby contributing to the ongoing discourse on provenance representation in the Digital Humanities.

## 1. Introduction

Recent years have seen the proliferation of numerous digital collections spanning various disciplinary fields under the expansive "big tent" (Svensson, 2012) of Digital Humanities. The data within them must be managed to be trustworthy, a goal usually achieved through the addition of provenance information, i.e., contextual metadata, primarily related to the identification of entities, such as the agent for the production, the date for the action and the reference sources (Gil et al., 2010). Moreover, in many humanities disciplines, the concept of "truth" is inherently tied to provenance, as truth is often considered a statement with adequate supporting sources. As if this were not enough, sources may be at odds with each other, and it is essential to keep track of such conjectures (Barabucci et al., 2022; Daquino et al., 2022).

However, storing provenance information alone is insufficient. Mechanisms to track changes in metadata related to cultural objects are crucial for maintaining data trustworthiness. Metadata evolves due to the natural progression of concepts or the correction of errors, and the latest version of knowledge may not always be the most accurate. Representing such dynamic information in RDF (Resource Description Framework) remains an open challenge. The foundational technologies of the Semantic Web—SPARQL (SPARQL Protocol and RDF Query Language), OWL (Web Ontology Language), and RDF—did not initially provide effective mechanisms to annotate statements with metadata information. This limitation has led to the introduction of numerous metadata representation models, none of which have become widely accepted standards for tracking both provenance and changes in RDF entities.

This article presents a comprehensive literature review on the representation models for provenance in RDF, specifically focusing on cultural heritage metadata. The goal is to provide an extensive overview of the existing models and approaches, facilitating an informed choice for practitioners and researchers in the field. We examine both the W3C standard proposals, such as RDF Reification and n-ary relations, and a broad array of other systems that have been developed over the years to address the shortcomings of these standards.

The structure of this article is as follows: Section 2 explores provenance for the Semantic Web, addressing the general challenge of provenance representation and examining the specific proposals of the W3C. Section 3 focuses on metadata representation models for provenance in RDF, while Section 4 discusses knowledge organization systems, specifically ontologies, for structuring provenance information. Section 5 provides a comprehensive discussion of the strengths and weaknesses of these models, considering their compliance with RDF standards, scalability, and applicability across different domains, and concludes the article by summarizing the main findings.

This review aims to contribute to the ongoing discourse on provenance representation by offering a structured analysis of existing approaches, thus supporting the development of more robust and scalable systems for managing cultural heritage metadata.

## 2. Foundations of provenance management in the Semantic Web

In his seminal work *Weaving the Web: the original design and ultimate destiny of the World Wide Web*, Tim Berners-Lee, the inventor of the World Wide Web, articulated a vision for a "Semantic Web" where computers would be capable of analyzing all data on the Web, encompassing content, links, and transactions between people and computers. This vision highlights the necessity of reliable data in a world where automatic data analysis systems manage critical aspects of trade, bureaucracy, and daily life (Berners-Lee, 1999). However, the open and inclusive nature of the Web presents a significant challenge: the prevalence of contradictory and questionable information. Therefore, to ensure data reliability, it is crucial to have metadata that provides context about the primary data source, the creator or modifier of the data, and the time of creation or modification. The foundational technologies of the Semantic Web—namely RDF, OWL, and SPARQL—were not originally designed to express such contextual information.

In 2010, the Provenance Incubator Group was established by the W3C to review the state of the art and develop a roadmap for provenance in Semantic Web technologies (Gil et al., 2010). One of the first challenges faced by the group was identifying a shared and universal definition of "provenance", a task that proved impossible given its broad and multisectoral nature. Consequently, a working definition was adopted, limited to the context of the Web: "Provenance of a resource is a record that describes entities and processes involved in producing and delivering or otherwise influencing that resource. Provenance provides a critical foundation for assessing authenticity, enabling trust, and allowing reproducibility. Provenance assertions are a form of contextual metadata and can themselves become important records with their own provenance".

Building on this definition, the group compiled 33 use cases to formulate scenarios and requirements, covering diverse fields such as eScience, eGovernment, business, manufacturing, cultural heritage, and library science. These use cases led to the development of three illustrative scenarios: a news aggregator, the study of an epidemic, and a business contract. The second scenario, the study of the epidemic, is particularly interesting for the case study of this work because it focuses on the reuse of scientific data. In this scenario, a fictitious epidemiologist named Alice is studying the spread of a new disease called owl flu. She needs to integrate structured and unstructured data from different sources, to understand how data has evolved through provenance and version information. In addition, she needs to justify the results obtained by supporting the validity of the sources used, reusing data published by others in a new context, and using the provenance to repeat previous analyses with new data. Introducing the problem with a concrete and complex example is helpful to understand how multifaceted and multidimensional it is.

Provenance can be evaluated under three broad categories: content, management, and usage —as defined in the study by Gil et al. (2010). These categories encompass various dimensions that are essential for understanding the requirements and challenges associated with provenance representation in the Semantic Web. Table 1 summarizes these dimensions.

*Table 1 Dimensions of provenance described and classified under three broad categories: content, management, and usage (adapted from Gil et al., 2010).*

| **Category** | **Dimension** | **Description** |
|---|---|---|
| **Content** | Object | The artifact that a provenance statement is about. |
| | Attribution | The sources or entities that contributed to creating the artifact in question. |
| | Process | The activities (or steps) that were carried out to generate or access the artifact at hand. |
| | Versioning | Records of changes to an artifact over time and what entities and processes were associated with those changes. |
| | Justification | Documentation recording why and how a particular decision is made. |
| | Entailment | Explanations showing how facts were derived from other facts. |
| **Management** | Publication | Making provenance available on the Web. |
| | Access | The ability to find the provenance for a particular artifact. |
| | Dissemination | Defining how provenance should be distributed and its access be controlled. |
| | Scale | Dealing with large amounts of provenance. |
| **Use** | Understanding | How to enable the end-user consumption of provenance. |
| | Interoperability | Combining provenance produced by multiple different systems. |
| | Comparison | Comparing artifacts through their provenance. |
| | Accountability | Using provenance to assign credit or blame. |
| | Trust | Using provenance to make trust judgments. |
| | Imperfections | Dealing with imperfections in provenance records. |
| | Debugging | Using provenance to detect bugs or failures of processes. |

Many data models, annotation frameworks, vocabularies, and ontologies have been introduced to meet the above requirements. A comprehensive overview of these strategies will be provided in the following sections, detailing their advantages and disadvantages.

In order to conduct the review, we adopted a citation-based strategy (Figure 1), also known as "snowballing" (Wohlin, 2014), which consists of exploding the bibliography from a seed paper (Lecy & Beatty, 2012), which was Sikos & Philp (2020).

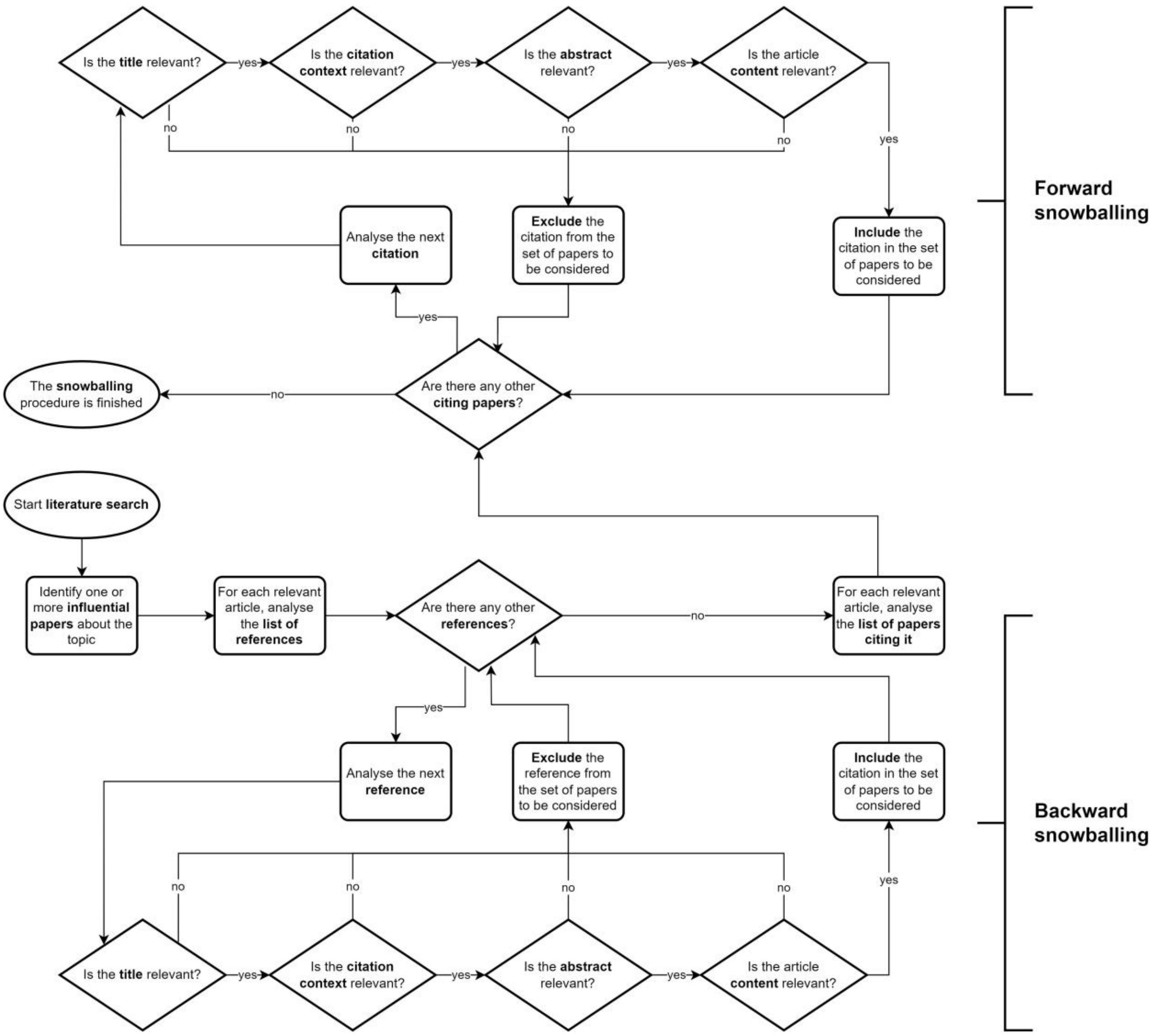

*Figure 1 The snowballing procedure adopted for conducting a systematic review.*

To address the challenge of representing provenance in RDF, the W3C has proposed several models, notably RDF Reification (Manola & Miller, 2004) and n-ary relations (W3C, 2006). RDF Reification is the earliest and most fundamental approach proposed by the W3C for annotating RDF statements. In RDF Reification, a statement (or triple) is converted into a resource, allowing metadata to be attached to it. This process involves creating four additional triples for each original statement, specifying the subject, predicate, object, and the statement itself.

An example of RDF Reification is provided in Listing 1, which illustrates how a simple RDF statement asserting the date of a manuscript is transformed into a set of reified triples. The new resource, representing the original statement, can then serve as the subject of additional provenance triples, such as those documenting the curator responsible for recording the information.

```
ex:manuscript10245 ex:hasDate “15th Century”.

ex:triple12345 rdf:type rdf:Statement ;

      rdf:subject ex:manuscript10245 ;

      rdf:predicate ex:hasDate ;

      rdf:object “15th Century” .

ex:triple12345 dc:creator ex:curator56789 .
```

*Listing 1 Representation of provenance using RDF Reification, where a triple is transformed into a statement resource to which metadata can be attached.*

Such methodology has a considerable disadvantage: the size of the dataset is at least quadrupled since subject, predicate, and object must be repeated to add at least one provenance’s information. There is a shorthand notation, the `rdf:ID` attribute in RDF/XML (Listing 2), but it is not present in other serializations.

```
<?xml version="1.0"?>

<!DOCTYPE rdf:RDF [<!ENTITY xsd "http://www.w3.org/2001/XMLSchema#">]>

<rdf:RDF

xmlns:rdf="http://www.w3.org/1999/02/22-rdf-syntax-ns#"
xmlns:dc="http://purl.org/dc/elements/1.1/"
xmlns:ex="http://www.example.com/terms/"
xml:base="http://www.example.com/2002/04/manuscripts">

    <rdf:Description rdf:ID="manuscript10245">

        <ex:hasDate rdf:ID="triple12345">

15th Century

  </ ex:hasDate>

    </rdf:Description>

    <rdf:Description rdf:about="#triple12345">

        <dc:creator rdf:resource="http://www.example.com/curatorid/56789"/>

    </rdf:Description>

</rdf:RDF>
```

*Listing 2 Generating reifications using `rdf:ID`.*

Finally, composing SPARQL queries to obtain provenance annotated through RDF Reification is cumbersome: to identify the URI of the reification, it is necessary to explicitly match the entire reference triple. Listing 3 provides an example of a SPARQL query that retrieves the curator responsible for recording the date of a manuscript.

```
PREFIX ex: <http://example.org/>
PREFIX rdf: <http://www.w3.org/1999/02/22-rdf-syntax-ns#>
PREFIX dc: <http://purl.org/dc/elements/1.1/>

SELECT ?curator WHERE {
  ?reifiedStatement rdf:type rdf:Statement ;
                    rdf:subject ex:manuscript10245 ;
                    rdf:predicate ex:hasDate ;
                    rdf:object "15th Century";
                    dc:creator ?curator .
}
```

*Listing 3 Retrieving provenance metadata in RDF reification using SPARQL.*

For all the mentioned reasons, there are several deprecation proposals for this syntax, including that by David Beckett, one of the editors of RDF in 2004, and RDF/XML (Revised) W3C Recommendation: "There are a few RDF model parts that should be deprecated (or removed if that seems possible), in particular reification which turned out not to be widely used, understood or implemented even in the RDF 2004 update" (Beckett, 2010).

After RDF Reification, in 2006, the W3C published a note that suggested a new approach to express provenance, called n-ary relations (W3C, 2006). In RDF and OWL, properties are always binary relationships between two URIs or a URI and a value. However, sometimes it is convenient to connect a URI to more than one other URI or value, such as expressing the provenance of a certain relationship.

The n-ary relations approach achieves this by introducing a new node, i.e. a blank node, that serves as an intermediary between the subject and the object, allowing additional metadata to be attached. Using the same example as before, a statement such as `ex:manuscript10245 ex:hasDate "15th Century"` can be transformed into an n-ary relation by introducing a blank node to encapsulate the provenance information, as shown in Listing 4. Here, a blank node (`_:date`) acts as an intermediary, linking the manuscript entity to both the date value and the provenance metadata.

From a structural perspective, n-ary relations share similarities with RDF Reification, but there is a fundamental difference in their approach. RDF Reification converts the entire triple into an entity that can be annotated with additional metadata, whereas n-ary relations introduce an intermediary node that encodes a reified relationship. This method avoids repeating all three elements of the original triple, as is required in RDF Reification, and instead only introduces a new node as the object of the original statement. However, a drawback of this approach is that the blank node introduced cannot be globally dereferenced, which can limit interoperability in certain contexts.

```
@prefix ex: <http://example.org/> .
@prefix dc: <http://purl.org/dc/elements/1.1/> .

ex:manuscript10245 ex:hasDate _:date .
_:date ex:hasDateValue "15th Century" ;
       dc:creator ex:curator56789 .
```

*Listing 4 Representation of provenance using the n-ary relation approach, where a blank node encapsulates both the date value and its associated provenance metadata.*

In summary, RDF Reification and n-ary relations are the only standard and alternative recommended by W3C to describe the provenance in RDF and have considerable design flaws. For this reason, different approaches have been proposed, as will be clarified in the following section. Among these, RDF-star (Gschwend & Lassila, 2022) has gained significant attention and is currently under consideration as a W3C recommendation. While still in the Working Draft stage, RDF-star addresses many of the issues inherent in traditional reification techniques and has been increasingly adopted in various implementations. A more detailed discussion of RDF-star is provided later in Section 3.

## 3. A comprehensive review of provenance representation systems

Since 2005, numerous approaches have been proposed to represent provenance in RDF datasets. In our classification, we adopt the framework proposed by Sikos & Philp (2020), which categorizes these approaches based on multiple dimensions: semantics, tuple typology (i.e., the number of elements in an RDF tuple, such as triples, quadruples, or quintuples), standard compliance, reliance on external vocabularies, blank node management, granularity, and scalability. These approaches can be broadly categorized into three main types: encapsulating provenance within RDF triples, associating provenance through RDF quadruples, and extending the RDF data model itself.

- Encapsulating provenance in RDF triples: n-ary relations (W3C, 2006), PaCE (Sahoo et al., 2010), singleton properties (Nguyen et al., 2014), Wikibase qualifiers (Vrandečić & Krötzsch, 2014), and the E13 Attribute in CIDOC CRM (Doerr, 2003).
- Associating provenance to the triple through RDF quadruples: named graphs (Carroll et al., 2005), RDF/S graphsets (Pediaditis et al., 2009), RDF triple coloring (Flouris et al., 2009), nanopublications (Gil et al., 2010), and conjectural graphs (Daquino et al., 2022).
- Extending the RDF data model: Notation3 Logic (Berners-Lee, 2005), RDF$^{+}$ (Dividino et al., 2009), SPOTL(X) (Hoffart et al., 2013), annotated RDF (aRDF) (Udrea et al., 2010); (Zimmermann et al., 2012), and RDF-star (Hartig & Thompson, 2019).

The **Provenance Context Entity (PaCE)** approach (Sahoo et al., 2010) proposes a method for attaching provenance to RDF data without using standard reification or blank nodes. Instead, each element (subject, predicate, object) can be "minted" as a separate URI that includes the source information, thereby embedding provenance directly in the resource names. In Listing 5, we illustrate how PaCE might look for the statement "manuscript 10245 has date '15th Century'" when the source is `ex:ArchiveA`.

```
@prefix provenir: <http://knoesis.wright.edu/provenir/provenir.owl#> .

ex:manuscript10245_ArchiveA a ex:manuscript123 .
ex:manuscript10245_ArchiveA ex:hasDate "15th Century" .
ex:manuscript10245_ArchiveA provenir:derives_from ex:ArchiveA .
```

*Listing 5 Representation of provenance using PaCE.*

This approach creates a new instance (`ex:manuscript10245_ArchiveA`) representing the original entity (`ex:manuscript10245`) within the context of its provenance source (`ex:ArchiveA`). The property `provenir:derives_from` from the Provenir Ontology (Sahoo & Sheth, 2009) was originally proposed to associate provenance metadata. The use of `provenir:derives_from` in PaCE introduces a meta-rule that extends RDF(S) semantics while maintaining compatibility with standard RDF 1.1. Specifically, when two RDF triples share the same value for `provenir:derives_from`, they are considered to belong to the same provenance context.

Unlike RDF reification, which requires multiple joins to reconstruct triples, PaCE offers a more compact representation. However, its adoption has been hindered by significant limitations. The most restrictive is that PaCE can only indicate the primary source of an entity, without supporting any additional provenance details.

While it would not be entirely accurate to say that PaCE was never implemented, as it was theoretically applied to the Biomedical Knowledge Repository (BKR) dataset, this implementation is effectively inaccessible. The BKR dataset itself is not available online, and the only reference to PaCE's use in a practical setting comes from a benchmark that evaluated different provenance-tracking models. However, even the benchmark data can now only be retrieved through the Internet Archive[1], and within those archived files, the actual datasets referenced in the study were originally hosted on Google Docs, which are no longer accessible.

Another approach that encapsulates provenance within triples is the **singleton property** method. Inspired by set theory—where a singleton set contains exactly one element—the singleton property approach creates "a unique property instance representing a newly established relationship between two existing entities in one particular context" (Nguyen et al., 2014). In practice, subjects are connected to objects using these unique properties, which are linked to a generic predicate via a dedicated predicate (commonly `rdf:singletonPropertyOf`). This unique property instance then serves as the carrier for additional provenance metadata.

The singleton property approach was introduced to address the limitations of PaCE, particularly its inability to represent provenance beyond identifying a primary source. While PaCE required minting new URIs that embedded provenance information, singleton properties instead generate unique predicates for each assertion, allowing metadata to be attached without altering the identity of the entities involved. This fundamental difference enables more flexible provenance tracking while maintaining compatibility with standard RDF structures.

[1] https://web.archive.org/web/20190421211038/http://wiki.knoesis.org/index.php/Singleton_Property#BKR

For example, a statement such as `ex: manuscript10245 ex:hasDate “15th Century”` can be transformed using singleton properties, where a unique property (`ex:hasDate#1`) is introduced to encapsulate provenance information. This transformation is illustrated in Listing 6, where the singleton property is explicitly linked to the generic predicate (`ex:hasDate`) and further annotated with provenance metadata (`dc:creator ex:curator56789`).

```
@prefix ex: <http://example.org/> .

@prefix dc: <http://purl.org/dc/elements/1.1/> .

@prefix rdf: <http://www.w3.org/1999/02/22-rdf-syntax-ns#>

ex:manuscript10245 ex:hasDate#1 "15th Century".

ex:hasDate#1 rdf:singletonPropertyOf ex:hasDate ;

           dc:creator ex:curator56789 .
```

*Listing 6 Representation of provenance using the singleton property approach, where a unique property instance (ex : `hasDate#1`) encapsulates additional metadata.*

While this strategy has demonstrated advantages in terms of reducing query size and improving query execution time when compared to PaCE (Nguyen et al., 2014), it also suffers from disadvantages in scenarios where multiple predications share the same source. In addition, singleton properties, like PaCE, rely on non-standard terms and face scalability challenges, yet they remain fully compliant with the RDF data model and SPARQL and are serializable in any RDF format.

In addition to these, the **Wikibase ontology**, which underlies Wikidata, presents another approach for encapsulating provenance within triples. In Wikibase, each statement is treated as a separate entity—commonly referred to as a statement node—designed to encapsulate the primary assertion along with its associated provenance metadata. For instance, consider the snippet from the Wikidata RDF dump for Codex Leicester (`wd:Q683814`) in Listing 7. In this example, the item "Codex Leicester" is connected to a statement via property `p:P127` ("owned by"). The statement itself, of type `wikibase:Statement` (and marked with `wikibase:BestRank`), contains the main assertion—indicating that Codex Leicester is owned by Bill Gates (`wd:Q5284`)—and a qualifier (`pq:P580`) that specifies the start time of this ownership.

```
@prefix wd: <http://www.wikidata.org/entity/> .
@prefix wdt: <http://www.wikidata.org/prop/direct/> .
@prefix p: <http://www.wikidata.org/prop/> .
@prefix ps: <http://www.wikidata.org/prop/statement/> .
@prefix pq: <http://www.wikidata.org/prop/qualifier/> .
@prefix wikibase: <http://wikiba.se/ontology#> .

wd:Q683814 p:P127 s:Q683814-d5054028-4345-88f5-c931-b20e3241efbb .
s:Q683814-d5054028-4345-88f5-c931-b20e3241efbb
    a wikibase:Statement,
      wikibase:BestRank ;
    ps:P127 wd:Q5284 ;
    pq:P580 "1994-01-01T00:00:00Z" .
```

*Listing 7 Wikibase RDF representation for Codex Leicester—demonstrating how the item (`wd:Q683814`) is linked via property `p:P127` to its ownership statement, which asserts that Bill Gates (`wd:Q5284`) is the owner, with a qualifier (`pq:P580`) indicating the start time of ownership.*

This design is fully RDF 1.1 compliant since it uses only triples and standard serialization formats (such as Turtle, RDF/XML, or JSON-LD), though it leverages a specialized Wikibase vocabulary.

Similarly, within the cultural heritage domain, **CIDOC CRM** employs its E13 Attribute Assignment class to record the action of making assertions about an object's property or the relationship between two items or concepts. Conceptually similar to the Wikibase mechanism, the E13 Attribute Assignment class permits the attachment of provenance metadata—such as the asserting agent, the date of assertion, and the contextual conditions—directly to the assertion. This approach not only captures the initial attribution (for example, recording that an artwork was produced by Rembrandt) but also supports the documentation of subsequent or even conflicting assertions over time, reflecting the evolving nature of cultural heritage interpretations.

Associating provenance through RDF quadruples includes named graphs (Carroll et al., 2005), RDF/S graphsets (Pediaditis et al., 2009), RDF triple coloring (Flouris et al., 2009), nanopublications (Groth et al., 2010), and conjectural graphs (Daquino et al., 2022).

**Named Graphs** extend the basic RDF model, traditionally composed of triples, by adding a fourth element to form quadruples. This fourth element is the graph URI, which serves as a contextual identifier. This structure allows RDF statements to not only describe resources but also to encapsulate metadata about the graphs themselves.

The serialization of Named Graphs can be achieved through extensions of RDF/XML, Turtle, and N-Triples, known respectively as TriX (Carroll & Stickler, 2004), TriG (W3C, 2024b), and N-Quads (W3C, 2024a) . These formats are standardized and compatible with SPARQL.

The W3C has proposed Named Graphs as a foundational element for expressing the provenance of scientific statements in a model known as "**nanopublications**." A nanopublication typically consists of three interrelated Named Graphs: one for the core data, one for provenance, and one for publication metadata. This structure provides a robust framework for ensuring the credibility and reproducibility of scientific information (Groth et al., 2010).

Named Graphs enhance RDF by allowing triples to be grouped into distinct graphs identified by URIs, aiding in managing provenance and trust. However, a limitation arises when handling implicit triples generated through inference mechanisms. For example, RDF Schema (W3C, 2014) introduces inference rules that generate implicit triples, which are not explicitly declared. When a Named Graph is deleted, the associated inference rules—and hence the derived triples—are also lost because Named Graphs do not inherently distinguish between explicit and inferred triples.

To overcome these limitations, RDF/S graphsets (Pediaditis et al., 2009) and RDF triple coloring (Flouris et al., 2009) extend Named Graphs' capabilities. **RDF/S graphsets** extend Named Graphs by allowing co-owned inferred triples to reside in a graphset URI rather than a single named graph. In the example in Listing 8, the triple `ex:manuscript10245 rdf:type ex:AncientText` results from combining two named graphs: `ex:namedGraph1`, which states that `ex:manuscript10245` is of type `ex:Manuscript` and that `ex:Manuscript` is a subclass of `ex:Document`, and `ex:namedGraph2`, which declares that `ex:Document` is a subclass of `ex:AncientText`. By applying transitive reasoning over `rdfs:subClassOf`, it follows that `ex:manuscript10245` is of type `ex:AncientText`. Since this conclusion does not originate from a single named graph but rather from their combination, it is explicitly recorded under `ex:graphset1` to preserve its provenance and co-ownership, ensuring that the inferred knowledge remains intact even if one contributing named graph is altered or deleted.

```
@prefix ex: <http://example.org/> .
@prefix rdf: <http://www.w3.org/1999/02/22-rdf-syntax-ns#> .
@prefix rdfs: <http://www.w3.org/2000/01/rdf-schema#> .

GRAPH ex:namedGraph1 {
   ex:manuscript10245 rdf:type ex:Manuscript .
   ex:Manuscript rdfs:subClassOf ex:Document .
}
GRAPH ex:namedGraph2 {
   ex:Document rdfs:subClassOf ex:AncientText .
}
GRAPH ex:graphset1 {
   ex:graphset1 rdf:type ex:Graphset ;
      ex:hasNamedGraph ex:namedGraph1, ex:namedGraph2 .
   ex:manuscript10245 rdf:type ex:AncientText .
}
```

*Listing 8 An RDF/S graphset example in TriG syntax, where the triple `ex:manuscript10245 rdf:type ex:AncientText` is recorded under `ex:graphset1` to indicate that it is inferred from both `ex:namedGraph1` and `ex:namedGraph2`.*

Please note that the TriG serialization presented here is a deduction based on the analysis of the master thesis works on RDF/S graphsets (Pediaditis, 2008) and its related paper (Pediaditis et al., 2009). To the best of our knowledge, no official serialization or concrete implementation exists, as the model has been described only in terms of logic and semantics. Similarly, no dedicated namespace for graphsets has been defined in the literature.

In contrast, **RDF Triple Coloring** provides a way to track the provenance of both explicit and inferred statements using standard RDF constructs. As with named graphs, each triple is stored as a quadruple `(s,p,o,c)`, where `c`, i.e., the “color”, is a URI denoting the source graph. However, when an implicit triple follows from two or more explicit statements, the inferred statement goes into a new color, conceptually “color1 + color2.” Listing 9 shows a TriG example. As for RDF/S graphsets, to the best of our knowledge, no RDF serialization for RDF triple coloring exists in the literature. The serialization provided in Listing 9 is therefore a deduction, based on the logical descriptions and algorithms found in Flouris et al. (2009).

```
@prefix ex:   <http://example.org/> .
@prefix rdfs: <http://www.w3.org/2000/01/rdf-schema#> .

ex:Color1 {
  ex:manuscript10245 rdf:type ex:Manuscript .
  ex:Manuscript rdfs:subClassOf ex:Document .
}
ex:Color2 {
  ex:Document rdfs:subClassOf ex:AncientText .
}
ex:Color1_2 {
  ex:manuscript10245 rdf:type ex:AncientText .
}
```

*Listing 9 Representation of explicit and inferred statements Using RDF Triple Coloring in TriG.*

Despite the presence of Named Graphs in both approaches, RDF triple coloring and RDF/S graphsets diverge in how they extend provenance tracking. In the triple-coloring model, each color is simply a Named Graph within the scope of the RDF 1.1 Recommendation, and it fully aligns with SPARQL 1.1's native support for GRAPH, FROM NAMED, and other relevant keywords. No modifications to the RDF data model or the SPARQL query language are required; the engine treats "color" URIs as any other graph identifier.

By contrast, RDF/S graphsets formalize an additional concept of "graphsets" that group named graphs, collectively assign ownership to their derived statements, and define custom entailment rules about how these sets interact. Although graphsets may use Named Graphs internally, they also impose extra metadata structures that do not map cleanly onto existing SPARQL constructs. Standard SPARQL 1.1 operations (e.g., SELECT, INSERT, GRAPH) cannot directly address a "graphset" as a single resource unless the engine is extended to interpret this extra layer of grouping. Moreover, the graphsets approach includes specialized semantics for deciding when a derived triple belongs to a particular set of named graphs, a notion outside RDF 1.1's default inference model. As a result, RDF/S graphsets necessitate non-standard processing, preventing off-the-shelf triplestores from handling them natively without custom extensions.

On the other hand, **Conjectural graphs** address the need to express uncertain or evolving claims in RDF. In the Arts, for instance, scholars often debate claims with no clear consensus. Conjectural graphs allow the representation of such uncertain claims without asserting their truth, using a unique predicate for each conjectured triple. This approach ensures that the original triples are not asserted, fulfilling the need to Express Without Asserting (EWA) (Daquino et al., 2022).

Consider the painting *Girl reading a letter at an open window*, which has been attributed to Rembrandt, Hooch, and Vermeer. Using conjectural graphs, we can represent these competing attributions without asserting any of them as absolute truth. Each competing attribution is represented by a unique predicate, ensuring that these statements are treated as conjectural forms rather than asserted facts. These unique predicates are linked to the original predicate to maintain clarity and traceability. Listing 10 shows an example using RDF and Turtle syntax.

```
@prefix ex: <http://example.org/> .
@prefix conj: <http://w3id.org/conjectures/> .
@prefix crm: <http://www.cidoc-crm.org/cidoc-crm/> .

ex:painting a crm:E22_Human-Made_Object .
ex:painting_activity a crm:E14_Activity ;
      crm:P108_has_produced ex:painting .

GRAPH <http://example.org/conjecture/rembrandt> {
  ex:painting_activity ex:cp1 ex:Rembrandt .
  ex:cp1 conj:isAConjecturalFormOf crm:P14_carried_out_by .
}

GRAPH <http://example.org/conjecture/hooch> {
  ex:painting_activity ex:cp2 ex:Hooch .
  ex:cp2 conj:isAConjecturalFormOf crm:P14_carried_out_by .
}

GRAPH <http://example.org/conjecture/vermeer> {
  ex:painting_activity ex:cp3 ex:Vermeer .
  ex:cp3 conj:isAConjecturalFormOf crm:P14_carried_out_by .
}
```

*Listing 10 Example of using conjectural graphs to represent conflicting attributions of a painting.*

In this example, each attribution (Rembrandt, Hooch, Vermeer) is represented by a unique conjectural predicate (`ex:cp1`, `ex:cp2`, `ex:cp3`). These conjectural predicates are mapped to the original predicate `crm:P14_carried_out_by` using the `conj:isAConjecturalFormOf` property. This mapping ensures that the original statements are not asserted, providing a clear distinction between conjectures and established facts.

Conversely, quadruples are not the only strategy to attach provenance information to RDF triples. Additionally, the RDF data model can be extended to achieve this goal. The first proposal of this kind was **Notation3 Logic**, which introduced the *formulae* (Berners-Lee, 2005). *Formulae* allow producing statements on N3 sentences, which are encapsulated by the syntax `{...}`. Berners-Lee and Connolly also proposed a *patch file format* for RDF deltas, or three new terms, using N3 (Berners-Lee & Connolly, 2004):

1. `diff:replacement`, that allows expressing any change. Deletions can be written as `{…} diff:replacement {}`, and additions as `{} diff:replacement {…}`.
2. `diff:deletion`, which is a shortcut to express deletions as `{…} diff:deletion {…}`.
3. `diff:insertion`, which is a shortcut to express additions as `{…} diff:insertion {…}`.

The main advantage of this representation is its economy: given two graphs G1 and G2, its storage cost is directly proportional to the difference between the two graphs. Therefore, it is a scalable approach. However, while conforming to the SPARQL algebra, N3 does not comply with the RDF data model and relies on the N3 Logic Vocabulary.

Adopting a completely different perspective, **RDF⁺** solves the problem by attaching a provenance property and its value to each triple, forming a quintuple (Table 2). In addition, it extends SPARQL with the expression `WITH META Metalist`, which includes graphs specified in `Metalist`, containing RDF⁺ meta knowledge statements (Dividino et al., 2009). To date, RDF⁺ is not compliant with any standard, neither the RDF data model, nor SPARQL, nor any serialization formats.

*Table 2 An RDF⁺ quintuple.*

| Subject | Predicate | Object | Meta-property | Meta-value |
|---|---|---|---|---|
| `ex: manuscript10245` | `ex:hasDate` | "15th Century" | `ex:accordingTo` | `ex:archive123` |

Also, **SPOTL(X)** allows expressing a triple provenance through quintuple (Hoffart et al., 2013). Indeed, the framework's name means Subject Predicate Object Time Location. Optionally, it is possible to create sextuples that add context to the previous elements. SPOTL(X) is concretely implemented in YAGO[2], a knowledge base automatically built from Wikipedia, given the need to specify which time, space, and context a specific statement is true. Outside of YAGO, SPOTL(X) does not follow either the RDF data model or the SPARQL algebra, and there is no standard serialization format.

Similarly, **annotated RDF (aRDF)** does not currently have any standardization. A triple annotation has the form `s p:λ o`, where λ is the annotation, always linked to the property (Udrea et al., 2010). **Annotated RDF Schema** perfects this pattern by annotating an entire triple and presenting a SPARQL extension to query annotations, called AnQL (Zimmermann et al., 2012). In addition, it specifies three application domains: the temporal, fuzzy, and provenance domains (Table 3).

*Table 3 Annotated RDF Schema application examples.*

| Domain | Annotated triple | Meaning |
|---|---|---|
| Temporal | (ex:manuscript10245, dct:created, "1450-01-01"): [1450] | Manuscript was created in the year 1450 |
| Provenance | (ex: manuscript10245, dct:creator, ex:author123): Archive | Manuscript was created by Author123 according to the Archive |

[2] https://yago-knowledge.org/

| Fuzzy | (ex:manuscript10245, dct:condition, ex:good): 0.8 | Manuscript is in good condition to a degree of 0.8 |
|---|---|---|
| Combined | (ex:manuscript10245, dct:creator, ex:author123): <[1450], 1, Archive> | Manuscript was created by Author123 in 1450, according to the Archive |

The most recent proposal for extending the RDF data model to handle provenance information is **RDF-star**, which embeds triples into triples as the subject or object (Hartig & Thompson, 2019). RDF-star aims to replace RDF Reification with less verbose and redundant semantics. As there is no serialization to represent this syntax, Turtle-star, an extension of Turtle, includes triples within << and >>. Similarly, SPARQL-star is an RDF-star-aware extension for SPARQL.

Later, RDF-star was proposed to allow statement-level annotations in RDF streams by extending RSP-QL to RSP-QL-star (Keskisärkkä et al., 2019). YAGO4 has adopted RDF-star to attach temporal information to its facts, expressing the temporal scope through `schema:startDate` and `schema:endDate` (Pellissier Tanon & Suchanek, 2019). For example, to express that Douglas Adams, *Hitchhiker's Guide to the Galaxy*'s author, lived in Santa Barbara until 2001 when he died, YAGO4 records <<DouglasAdams schema:homeLocation SantaBarbara>> schema:endDate 2001 (Pellissier Tanon et al., 2020).

Despite being incompatible with RDF 1.1, it is worth mentioning that a W3C working group has recently published the first draft to make RDF-star a standard (Gschwend & Lassila, 2022). This move suggests that RDF-star is likely to become the new de facto standard for managing provenance in RDF, addressing the current limitations and providing a more efficient and comprehensive solution for handling metadata and inferred triples.

## 4. Knowledge Organisation Systems for RDF provenance

A wide range of ontologies and vocabularies represent provenance information, either upper ontologies, domain ontologies, or provenance-related ontologies.

- Upper ontologies: Proof Markup Language (da Silva et al., 2006), Provenance Ontology (Gil et al., 2010), and the Open Provenance Model (Moreau et al., 2011).
- Domain ontologies: SWAN Ontology for Neuromedicine (Ciccarese et al., 2008), Provenir Ontology for eScience (Sahoo & Sheth, 2009), and PREMIS for archived digital objects (Caplan, 2017).
- Provenance-related ontologies: Dublin Core Metadata Terms (DCMI Usage Board, 2020), and the OpenCitations Data Model (Daquino et al., 2020).

Among the upper ontologies, the **Open Provenance Model** stands out because of its interoperability. It describes the history of an entity in terms of processes, artifacts, and agents (Moreau et al., 2011), a pattern that will be discussed later about the PROV Data Model (Moreau & Missier, 2013). On the other hand, the **Proof Markup Language (PML)** is an ontology designed to support trust mechanisms between heterogeneous web services (da Silva et al., 2006).

Among the domain-relevant models, several ontologies have been developed to address the specific needs of their respective fields. For example, the **Provenir Ontology** is tailored to capture the dynamic processes and data evolution in scientific research (Sahoo & Sheth, 2009), and it has already been mentioned in Section 3 in relation to PaCE; **PREMIS** is designed to document the provenance of archived digital objects such as files and bitstreams (Caplan, 2017); and **Semantic Web Applications in Neuromedicine (SWAN)** Ontology supports the modeling of scientific discourse in the biomedical research context (Ciccarese et al., 2008).

About domain-relevant models, there is Finally, the **Dublin Core Metadata** Terms allows to express the provenance of a resource and specify what is described (e.g., `dct:BibliographicResource`), who was involved (e.g., `dct:Agent`), when the changes occurred (e.g., `dct:dateAccepted`), and the derivation (e.g., `dct:references`) (DCMI Usage Board, 2020).

All the requirements and ontologies mentioned have been merged into a single data model, the PROV Data Model (Moreau & Missier, 2013), translated into the **PROV Ontology** using the OWL 2 Web Ontology Language (Lebo et al., 2013). It provides several classes, properties, and restrictions, representing provenance information in different systems and contexts. Its level of genericity is such that it is even possible to create new classes and data model-compatible properties for new applications and domains. Just like the Open Provenance Model, PROV-DM captures the provenance under three complementary perspectives:

- *Agent-centered provenance* entails people, organizations, software, inanimate objects, or other entities involved in generating, manipulating, or influencing a resource. For example, it is possible to distinguish between the author, the editor, and the publisher concerning a journal article. PROV-O maps the responsible agent with `prov:Agent`, the relationship between an activity and the agent with `prov:wasAssociatedWith`, and an entity's attribution to an agent with `prov:wasAttributedTo`.

- *Object-centred-provenance*, which is the origin of a document's portion from other documents. Taking the example of the article, a fragment of it can quote an external document. PROV-O maps a resource with `prov:Entity`, whether physical, digital, or conceptual, while the predicate `prov:wasDerivedFrom` expresses a derivation relationship.
- *Process-centered provenance*, or the actions and processes necessary to generate a resource. For example, an editor can edit an article to correct spelling errors using the previous version of the document. PROV-O expresses the concept of action with `prov:Activity`, the creation of an entity with the predicate `prov:wasGeneratedBy`, and the use of another entity to complete a passage with `prov:used`.

The Graffoo diagram in Figure 2 (Falco et al., 2014) provides a high-level view of the discussed concepts' structure, constituting the so-called "starting point terms". PROV-O is more extensive and provides modularly sophisticated entities, agents, activities, and relationships, namely "expanded terms" and "qualified terms".

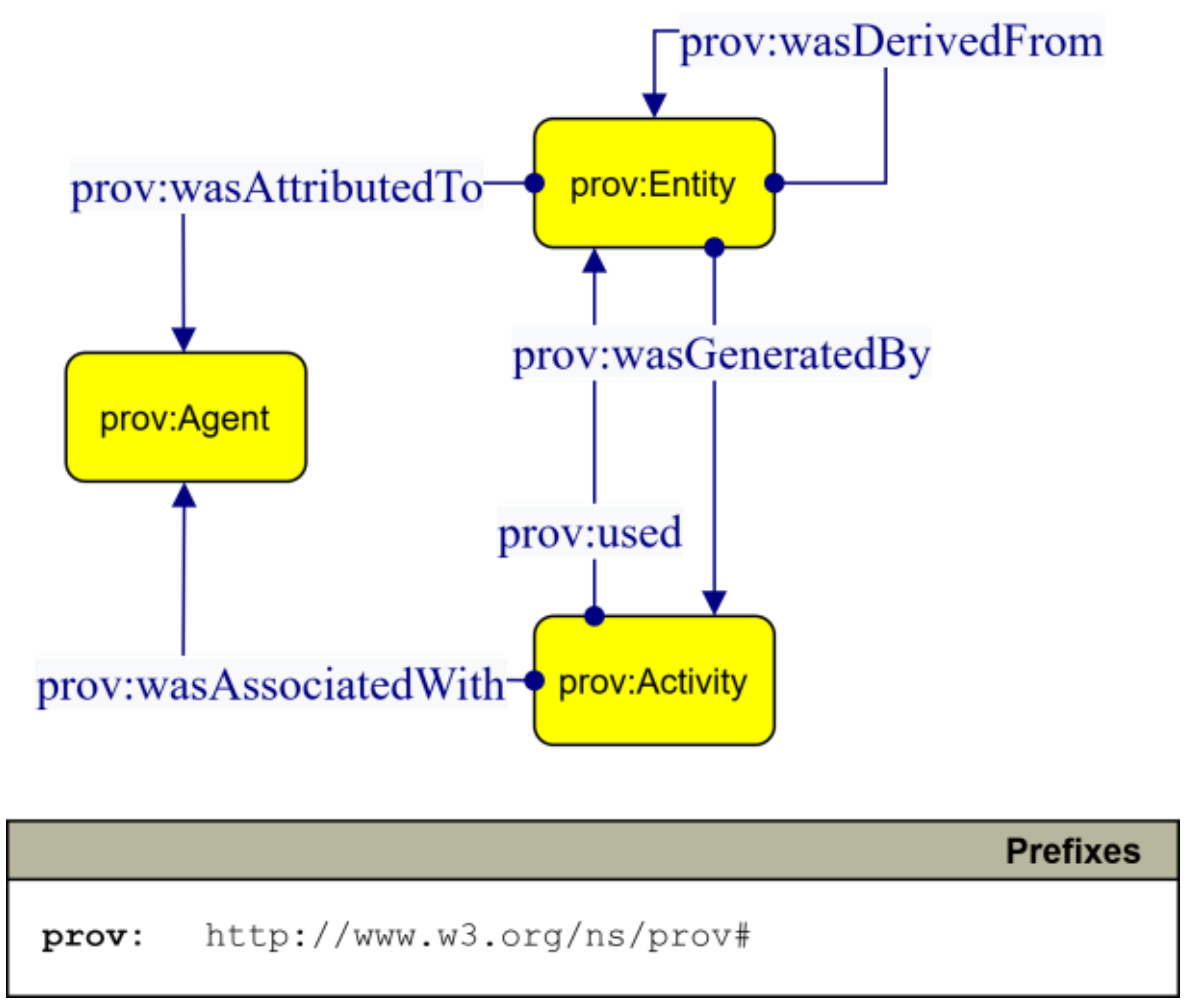

*Figure 2 High level overview of PROV records.*

Finally, The **OpenCitations Data Model (OCDM)** (Daquino et al., 2020) represents provenance and tracks changes in a way that complies with RDF 1.1. It relies on well-known and widely adopted standards such as PROV-O (Lebo et al., 2013), named graphs, and Dublin Core (Daquino et al., 2020). Each entity described by the OCDM is annotated with one or more snapshots of provenance. The snapshots are of type `prov:Entity` and are connected to the bibliographic entity described through `prov:specializationOf`, a predicate present in the mentioned "expanded terms". Being the specialization of another entity means sharing every aspect of the latter and, in addition, presenting more specific aspects, such as an abstraction, a context, or, in this case, a time. In addition, each snapshot records the validity dates (`prov:generatedAtTime`, `prov:invalidatedAtTime`), the agents responsible for both creation and modification of the metadata (`prov:wasAttributedTo`), the primary sources (`prov:hadPrimarySource`) and a link to the previous snapshot in time (`prov:wasDerivedFrom`). The model is summarised in Figure 3.

Furthermore, OCDM extends the Provenance Ontology by introducing a new property called `hasUpdateQuery`, a mechanism to record additions and deletions from an RDF graph with a SPARQL INSERT and SPARQL DELETE query string. The *snapshot-oriented* structure, combined with a system to explicitly indicate how a previous snapshot was modified to reach the current state, makes it easier to recover the current statements of an entity and restore an entity to a specific snapshot. The current statements are those available in the present dataset, while recovering a snapshot $s_i$ means applying the reverse operations of all update queries from $s_n$ to $s_{i+1}$ (Peroni et al., 2016).

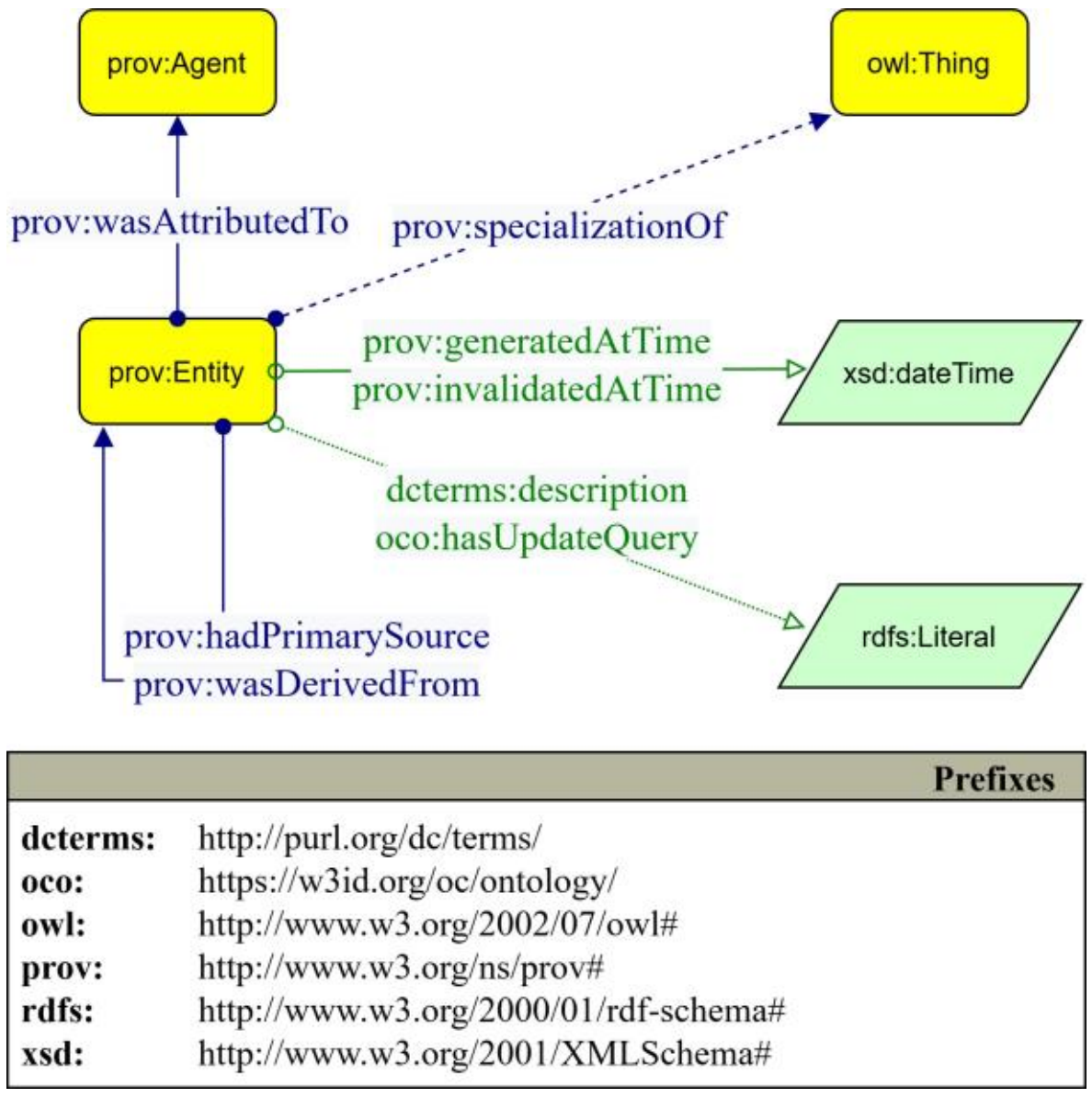

*Figure 3 Provenance in the OpenCitations Data Model.*

To conclude, some readers may wonder why this survey treats Wikibase qualifiers and CIDOC CRM (E13 Attribute Assignment) as mechanisms for encapsulating provenance rather than as ontologies in their own right. Although both are technically ontological frameworks, in this context they are noteworthy not for the specific ontology they offer, but for the reification strategy they implement. In other words, the real contribution of Wikibase qualifiers and CIDOC CRM E13 lies in how they attach provenance metadata directly to assertions within RDF triples (or their reified equivalents), rather than in providing a language for provenance representation per se. Consequently, they frequently defer to external provenance ontologies—such as PROV-O or Dublin Core—for expressing the provenance details themselves, underscoring that their primary innovation is the mechanism of embedding provenance data, rather than a standalone ontology for provenance.

## 5. Discussion and conclusion

Understanding the various models for representing provenance in RDF is crucial for making informed choices about which to implement, especially in the realm of Digital Humanities and cultural heritage metadata management. In this section, we will analyze the pros and cons of these models as summarized in Table 4 and illustrated in Figure 4, to identify the most effective current solutions.

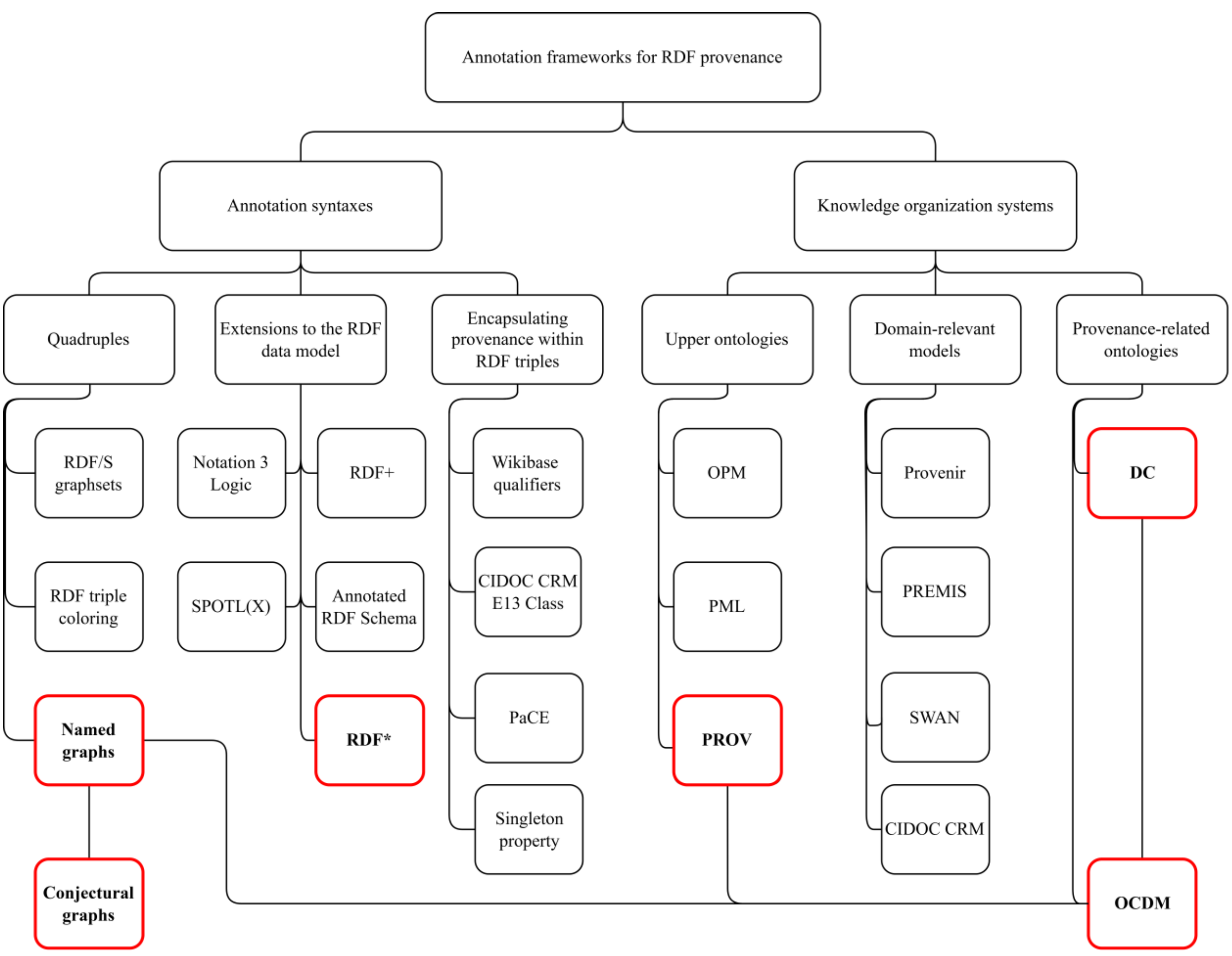

*Figure 4 Annotation frameworks for RDF provenance. The edges represent a membership relationship, while the bold type and red circling highlight the syntaxes we identified as the most relevant.*

*Table 4 Advantages and disadvantages of metadata representations models for RDF. A glossary of the acronyms can be consulted on Zenodo (Massari, 2023). The table was expanded from the one in (Sikos & Philp, 2020).*

| **Approach** | **Tuple type** | **Compliance with RDF** | **Compliance with SPARQL** | **RDF serialisations** | **External vocabulary** | **Scalable** | **Expressing without asserting** |
|---|---|---|---|---|---|---|---|
| Named graphs | Quadruple | yes | yes | TriG, TriX, N Quads | no | yes | Depends on the implementation |
| RDF/S graphsets | Quadruple | no | no | TriG, TriX, N Quads | no | yes | Depends on the implementation |
| RDF triple coloring | Quadruple | yes | yes | TriG, TriX, N Quads | no | yes | Depends on the implementation |
| N3Logic | Triple (in N3) | no | yes | N3 | N3 Logic Vocabulary | yes | yes |
| aRDF & Annotated RDF Schema | Non-standard | no | no | no | no | yes | yes |
| RDF$^+$ | Quintuple | no | no | no | no | yes | yes |
| SPOTL(X) | Quintuple/ sextuple | no | no | no | no | Depends on the implementation | yes |
| RDF-star | Non-standard | no | no | Turtle-star (non-standard) | no | yes | yes |
| Wikibase qualifiers | Triple | yes | yes | Turtle, N-Triples, RDF-JSON, JSON-LD, RDFa, HTML5 Microdata | Wikibase ontology | no | no |
| CIDOC CRM E13 Class | Triple | yes | yes | Turtle, N-Triples, RDF-JSON, JSON-LD, RDFa, HTML5 Microdata | CIDOC CRM ontology | no | no |
| PaCE | Triple | yes | yes | Turtle, N-Triples, RDF-JSON, JSON-LD, RDFa, HTML5 Microdata | Provenir ontology | no | no |
| Singleton property | Triple | yes | yes | RDF/XML, N3, Turtle, N-Triples, RDF-JSON, JSON-LD, RDFa, HTML5 Microdata | Singleton property | no | no |
| Conjectural graph | Quadruple | yes | yes | TriG, TriX, N Quads | Conjectural property | no | yes |

The analysis shows that Named Graphs, RDF-star, PROV-O, Dublin Core, Conjectural Graphs, and the OpenCitations Data Model stand out as the most effective models. Named Graphs use quadruples and are fully compliant with RDF and SPARQL standards. They support multiple serialization formats, such as TriG, TriX, and N-Quads, and are scalable, which makes them highly suitable for various applications including digital humanities and scholarly data.

RDF-star addresses verbosity and redundancy issues found in RDF reification by embedding triples within triples. Despite being non-standard, RDF-star is gaining acceptance due to its efficiency and compatibility with RDF data models and SPARQL. Its serialization format, Turtle-star, and practical applications, such as in YAGO4, demonstrate its potential for broader adoption.

The Provenance Ontology (PROV-O) provides a comprehensive framework for representing provenance across different contexts. As a W3C recommendation, PROV-O integrates a rich set of classes and properties, making it adaptable and robust for detailed provenance tracking.

Dublin Core (DC) is simple and widely adopted for basic metadata needs, especially in bibliographic contexts. Its compliance with RDF standards and ease of integration with other models, such as PROV-O and Named Graphs, makes it a foundational tool for metadata representation.

The OpenCitations Data Model (OCDM) extends the capabilities of these models by incorporating standards like PROV-O, named graphs, and Dublin Core to track changes and ensure data reliability. OCDM captures provenance snapshots, including validity dates, responsible agents, primary sources, and SPARQL update queries, making it a robust solution for managing evolving data.

Conjectural Graphs are particularly useful in domains like the arts and humanities, where representing uncertain or evolving claims without asserting them as facts is essential. They build on Named Graphs to handle conjectures effectively, thus offering a nuanced approach to metadata representation.

Other models, such as Singleton Properties and Wikibase qualifiers, offer innovative solutions to encapsulate provenance within RDF triples, reducing verbosity and improving query performance. However, these models face limitations due to their reliance on non-standard terms and scalability issues. Similarly, RDF/S graphsets, N3Logic, and $RDF^{+}$ struggle with standard compliance and integration, which hinders their broader adoption.

From the perspective of expressing statements without asserting them, different models exhibit varying levels of support. Named Graphs present an ambiguous case: while they allow grouping of statements under a graph URI, their semantics depend on the interpretation of the default graph and the conventions adopted by specific implementations. This ambiguity makes them a flexible but not inherently reliable method for expressing without asserting. Extensions such as RDF/S graphsets and RDF triple coloring add further granularity, allowing for the classification of statements into different levels of truth or certainty, but they still require additional interpretative mechanisms to prevent unwanted assertions.

N3Logic offers a more explicit approach to expressing without asserting by allowing graphs to be quoted rather than asserted. This mechanism enables the representation of statements within logical constructs without incorporating them into the asserted knowledge base, making it one of the most expressive solutions for non-asserted RDF representation. However, its adoption remains limited due to its reliance on Notation3 (N3), which is not natively supported in standard RDF tools.

RDF-star provides a lightweight syntax for embedding statements within other triples, and its quoting feature offers a way to refer to statements without asserting them. Similar functionality is observed in SPOTL(X) and $RDF^{+}$, which extend RDF to include additional elements such as context and temporal dimensions. Annotated RDF follows a similar logic. These models share the advantage of reducing verbosity compared to traditional reification, while still allowing for the expression of uncertainty or disputed claims. However, among them, only RDF-star is moving toward standardization, with increasing adoption in RDF tools and query languages like SPARQL-star. In contrast, SPOTL(X), RDF$^+$, and Annotated RDF, as said, remain confined to specific implementations, limiting their broader applicability.

By contrast, Wikibase qualifiers and CIDOC CRM E13 provide mechanisms for enriching statements with additional metadata but do not inherently prevent the base triple from being asserted. Likewise, PaCE and the Singleton Property method create new URIs for contextualizing statements, but the modified triples remain part of the asserted dataset. Conjectural Graphs, in contrast, explicitly structure statements as conjectural rather than asserted by introducing new predicate URIs that isolate the conjectured statement from the dataset's asserted portion. This allows competing or uncertain claims to be represented without treating them as established facts, making them particularly useful in the context of Digital Humanities and cultural heritage research.

In conclusion, the choice of a data model for capturing provenance in cultural heritage metadata is not a one-size-fits-all solution. It must balance practical considerations such as the volume and variability of the data, the complexity of uncertain or even contested knowledge, and the need for interoperability by complying with existing standards. Models like Named Graphs, Conjectural Graphs, RDF-star, and ontologies like PROV-O, Dublin Core and the OpenCitations Data Model offer concrete pathways to ensure robust, accurate, and transparent data management over time. However, the most effective approach will ultimately depend on how well it integrates with institutional workflows and addresses the specific challenges of representing provenance within diverse and evolving collections.

## Acknowledgements

This work has been partially funded by Project PE 0000020 CHANGES - CUP B53C22003780006, NRP Mission 4 Component 2 Investment 1.3, Funded by the European Union - NextGenerationEU.